\documentclass[aps,prl,reprint,twocolumn,amsmath,amssymb,showpacs,superscriptaddress]{revtex4-1}

\usepackage{color}
\usepackage{textcomp} 
\usepackage{mathrsfs,amsmath}
\usepackage{graphicx}
\usepackage{lipsum}
\usepackage{dcolumn}
\usepackage[mathlines]{lineno}
\usepackage{hyperref}
\usepackage{xcolor}
\hypersetup{
    colorlinks,
    linkcolor={red!50!black},
    citecolor={blue!50!black},
    urlcolor={blue!80!black}
}
\usepackage{bm}
\usepackage{epstopdf}
\usepackage{soul}
\usepackage{epsfig}
\usepackage{bbold}
\usepackage{braket}
\usepackage{physics}
\usepackage{gensymb}
\usepackage{amsmath,amssymb}

\begin{document}

\title{On the orbit-induced spin density of tightly focused optical vortex beams: ellipticity and helicity}
\author{Kayn A. Forbes}

\email{K.Forbes@uea.ac.uk}
\affiliation{School of Chemistry, University of East Anglia, Norwich Research Park, Norwich NR4 7TJ, United Kingdom}

\begin{abstract}
 It has recently been established that a linearly-polarized optical vortex possesses spin angular momentum density in the direction of propagation (longitudinal spin) under tight-focusing. The helicity of light has long been associated with longitudinal spin angular momentum. Here we show that the longitudinal spin density of linearly-polarized vortices is anomalous because it has no associated helicity. It was also recently determined that the polarization-independent helicity of tightly-focused optical vortices is associated with their transverse spin momentum density.  The key finding of this work is the fact that, in general, longitudinal spin can not necessarily be associated with helicity, and transverse spin is in general not associated with a zero helicity, and such extraordinary behaviour manifests most clearly for optical vortices under non-paraxial conditions.  
\end{abstract}

\maketitle

\section{Introduction}

The optical properties (energy, momentum, angular momentum, and helicity) of paraxial beams of light in general align with those of electromagnetic plane-waves. An important exception is the orbital angular momentum (OAM) carried by paraxial optical vortex beams along the direction of propagation \cite{andrews2012angular, shen2019optical, forbes2021structured}. Under non-paraxial conditions, i.e. tight-focusing, the electric and magnetic fields polarized in the direction of propagation (longitudinal) grow in magnitude relative to the transverse components \cite{novotny2012principles}, leading to a range of extraordinary behaviours for tightly-focused laser beams. Examples of these novel optical properties include transverse spin angular momentum \cite{aiello2015transverse, bliokh2015transverse, neugebauer2015measuring, neugebauer2018magnetic} and the optical helicity (also referred to as chirality) of vortex beams \cite{forbes2021measures, green2023optical, forbes2023customized}. One of the most surprising results, particularly given the conventional wisdom of the optical properties of light, is that input laser beams which are unpolarized before being tightly focused may possess both spin angular momentum \cite{eismann2021transverse} and helicity \cite{forbes2022optical} densities in the focal plane.

Very recently another remarkable property of non-paraxial laser beams was discovered: linearly-polarized optical vortex beams when tightly-focused possesses spin angular momentum density along the direction of propagation. A paraxial linearly polarized optical vortex beam before focusing has essentially zero spin angular momentum density, but in the focal plane after traversing a high NA lens a non-zero spin angular momentum density along the direction of propagation is generated. This property is unique to optical vortex beams, and there have been a number of studies to date on both scalar vortex beams \cite{yu2018orbit, kotlyar2020spin, forbes2021measures, zhang2022understanding} and vector vortex beams \cite{han2018catalystlike, li2018orbit, shi2018structured, meng2019angular, li2021spin, geng2021orbit, fang2021photoelectronic, zhang2022ultrafast, man2022polarization, forbes2024spin, liu2024manipulation}. The underlying mechanism is a form of spin-orbit interaction (SOI) of light \cite{bliokh2015spin}, but in contrast to the well-established spin-to-orbital angular momentum conversion it is an orbital to spin angular momentum density conversion. 

It is generally assumed that electromagnetic fields with longitudinal spin also possess helicity, and that the transverse spin of light is not associated with helicity. Indeed, a number of recent works have elaborated on the decomposition of longitudinal and transverse spin momentum density \cite{shi2022spin, shi2023dynamical, vernon2023decomposition}. Our aim in this work is to explicitly highlight the fact that non-paraxial optical vortex beams do not adhere to this conventional behaviour and in general optical helicity, polarization ellipticity, and spin momentum are not an interdependent trinity. 

\section{Analytical theory of a focused optical vortex Bessel beam}

A popular method to describe the focal fields of a non-paraxial beam is to use numerical integration techniques, e.g. Richards-Wolf diffraction theory \cite{novotny2012principles}. In this work we use an analytical approach based on pure Bessel modes due to their analytical simplicity and the fact they are solutions to both the paraxial and nonparaxial wave equations. A discussion of the different analytical and numerical techniques can be found in Ref \cite{peatross2017vector}. The analytical methods we favor here lead to simple analytical results and a deep insight into the novel contributions from specific higher-order field components to properties of electromagnetic fields. The electric field for a monochromatic scalar Bessel beam up to second-order in the paraxial parameter $k_t/k_z$ is \cite{forbes2024spin}

\begin{widetext}
\begin{align}
\mathbf{E} &= \Bigl[ J_{|{\ell}|}\text{e}^{i\ell\phi}(\alpha\mathbf{\hat{x}}+\beta\mathbf{\hat{y}}) \nonumber + \mathbf{\hat{z}}\frac{ik_t}{2k_z}\bigl((\alpha\pm i \beta)J_{|\ell|-1}\text{e}^{i(\ell\mp1)\phi} +(\pm i \beta-\alpha)J_{|\ell|+1}\text{e}^{i(\ell\pm1)\phi}\bigl) \nonumber + \mathbf{\hat{x}}\frac{k_t^2}{4k^2}\Bigl(2\alpha J_{|\ell|}\text{e}^{i\ell\phi} \nonumber \\
&+ J_{|\ell|-2}(\alpha \pm i \beta)\text{e}^{i(\ell\mp2)\phi} + J_{|\ell|+2}(\alpha \mp i \beta)\text{e}^{i(\ell\pm2)\phi}\Bigr) + \mathbf{\hat{y}}\frac{k_t^2}{4k^2}\Bigl(2\beta J_{|\ell|}\text{e}^{i\ell\phi} \nonumber + J_{|\ell|-2}(\pm i \alpha - \beta)\text{e}^{i(\ell\mp2)\phi} \nonumber \\
&+ J_{|\ell|+2}(\mp i \alpha -\beta)\text{e}^{i(\ell\pm2)\phi}\Bigr)\Bigr]\text{e}^{ik_z z}, 
\label{eq:1}
\end{align}
\end{widetext}

where $J_{|\ell|}[k_t r]$ is a Bessel function of the first-kind of order $|\ell|$ and argument $k_t r$ (the argument is suppressed in Eq. \eqref{eq:1} and throughout the manuscript for notational brevity, further we subsume units of electric field into the Bessel function); $\ell \in \mathbb{Z}$ is the topological charge, $\ell >0$ left-handed helical wavefronts, $\ell < 0$  right-handed helical wavefronts; $\phi$ is the azimuthal angle; $\alpha$ and $\beta$ are the Jones vector coefficients which are in general complex and $|\alpha|^2+|\beta|^2=1$; $k_z=\sqrt{k^2-k_t^2}$ is the longitudinal wavenumber and $k_t=\sqrt{k_x^2+k_y^2}$ the transverse wavenumber. The rule determining which sign to take for the $\pm$ and $\mp$ parts in Eq.~\eqref{eq:1} (and \eqref{eq:2}) is that if the topological charge of the mode is $\ell>0$ the upper-sign is taken; if $\ell<0$ the lower sign is taken. 

In language first introduced by Lax et al. \cite{lax1975maxwell},  Eq. \eqref{eq:1} contains the zeroth-order transverse $\text{T}_0$ (with respect to the smallness parameter $k_t/k_z)$, first-order longitudinal $\text{L}_1$, and second-order transverse field components $\text{T}_2$. The zeroth-order term in Eq. \eqref{eq:1} is the dominating term for a paraxial (well-collimated) Bessel beam; the ratio of $k_t/k_z$ becomes larger upon increasing the tightness of the focus eventually leading to the higher-order field components - first-order longitudinal and second-order transverse - becoming significant enough in magnitude compared to the zeroth-order fields to yield physically observable effects \cite{forbes2021relevance, green2023optical}. Before reaching the focusing lens, the beam is assumed well-collimated and paraxial: the only significant field components are the zeroth-order $\text{T}_0$ and the beam is described as being 2D-polarized in $x,y$. This polarization state of paraxial light is readily described by the well-known Poincar\'e sphere. In the focal plane of a tight focus, however, the higher-order fields $\text{L}_1$ and $\text{T}_2$ become significant in magnitude and in general lead to an extraordinarly complex polarization state. Although complex, in general the polarization state is now three-dimensional ($x,y,z$) and we speak of `3D-polarized' light \cite{alonso2023geometric}. Nonetheless, it must be noted that tightly focused light is routinely described in the literature as `linearly-polarized' or `circularly-polarized', for example, when what is technically meant is that these beams are 2D-linearly polarized or 2D-circularly-polarized, respectively, in the $x,y$-plane for a $z$-propagating beam (i.e. their paraxial sate of polarization before focusing). 

Maxwell's equations in free-space are dual symmetric and unlike paraxial beams, electromagnetic democracy is required in the expressions to describe the optical properties of non-paraxial beams \cite{bliokh2013dual}. Thus we therefore require the corresponding magnetic field of a Bessel beam: 

\begin{widetext}
\begin{align}
\mathbf{B} &= \Bigl[J_{|{\ell}|}\text{e}^{i\ell\phi} \frac{k_z}{k}(\alpha\mathbf{\hat{y}}-\beta\mathbf{\hat{x}}) + \mathbf{\hat{z}}\frac{ik_t}{2k}\bigl((\pm i\alpha - \beta)J_{|\ell|-1}\text{e}^{i(\ell\mp1)\phi} +(\pm i \alpha + \beta)J_{|\ell|+1}\text{e}^{i(\ell\pm1)\phi}\bigl) + \mathbf{\hat{x}}\frac{k_t^2}{4kk_z}\Bigl(-2\beta J_{|\ell|}\text{e}^{i\ell\phi} \nonumber \\
&+ J_{|\ell|-2}(\pm i \alpha - \beta)\text{e}^{i(\ell\mp2)\phi} + J_{|\ell|+2}(\mp i \alpha - \beta)\text{e}^{i(\ell\pm2)\phi}\Bigr) + \mathbf{\hat{y}}\frac{k_t^2}{4kk_z}\Bigl(2\alpha J_{|\ell|}\text{e}^{i\ell\phi} + J_{|\ell|-2}(\mp i \beta - \alpha)\text{e}^{i(\ell\mp2)\phi} \nonumber \\ 
&+ J_{|\ell|+2}(\pm i \beta -\alpha)\text{e}^{i(\ell\pm2)\phi}\Bigr)\Bigr]\frac{1}{c}\text{e}^{ik_z z}.
\label{eq:2}
\end{align}
\end{widetext}

The analytical electromagnetic fields Eqs. \eqref{eq:1} and \eqref{eq:2} containing field components up to second-order in the smallness parameter are used in this manuscript to describe a focused optical vortex beam.

\section{Spin Angular Momentum Density}

The spin angular momentum of light can be both longitudinal and transverse with respect to the direction of propagation \cite{andrews2012angular,bliokh2015transverse,aiello2015transverse}. Longitudinal spin angular momentum is much more familiar, and the spin of $\sigma \hbar \mathbf{\hat{z}}$ per photon for a $z$-propagating circularly polarized plane wave with helicity $\sigma = \pm 1$ per photon is a well-known result (and also highlights how the helicity and spin are entwined in paraxial fields). The cycle-averaged dual-symmetric spin momentum density $\mathbf{s}$ for a monochromatic beam is calculated using \cite{bliokh2013dual}

\begin{align}
\mathbf{s} = \frac{\epsilon_0}{4\omega} \text{Im}(\mathbf{E}^*\times\mathbf{E} + c^{2}\mathbf{B}^*\times\mathbf{B}). 
\label{eq:3}
\end{align}

The longitudinal spin angular momentum $s_z$ is generated by the cross product between the transverse $(x,y)$ polarized fields, such that up to second-order in the smallness parameter $s_{z}^\text{E} = \mathbf{E}_{\text{T}0} \times \mathbf{E}_{\text{T}0} + \mathbf{E}_{\text{T}0} \times \mathbf{E}_{\text{T}2} $ and $s_{z}^\text{B} = \mathbf{B}_{\text{T}0} \times \mathbf{B}_{\text{T}0} + \mathbf{B}_{\text{T}0} \times \mathbf{B}_{\text{T}2} $. The transverse spin of light manifests through the cross product of the transverse field with the $z$-polarized longitudinal component: $s_{x,y}^\text{E} = \mathbf{E}_{\text{T}0} \times \mathbf{E}_{\text{L}1}$ and $s_{x,y}^\text{B} = \mathbf{B}_{\text{T}0} \times \mathbf{B}_{\text{L}1}$, see Appendix for further details on the transverse spin. In this work we are specifically interested in the spin angular momentum generated by the cross terms $\mathbf{E}_{\text{T}0} \times \mathbf{E}_{\text{T}2}$ and $\mathbf{B}_{\text{T}0} \cross \mathbf{B}_{\text{T}2}$ for 2D linearly-polarized fields. 

Inserting Eqs.~\eqref{eq:1} and \eqref{eq:2} into Eq.~\eqref{eq:3} and assuming real $\alpha$ and $\beta$, i.e. a 2D-linearly polarized input beam gives

\begin{align}
s_{z}^\text{E} &= \frac{\epsilon_0 k_{t}^2}{2k^2\omega}\Big[J_{|\ell|}J_{|\ell|-2}\Big(\pm2\alpha\beta\sin2\phi\pm(\alpha^2-\beta^2)\cos2\phi\Big) \nonumber \\ & + J_{|\ell|}J_{|\ell|+2}\Big(\mp2\alpha\beta\sin2\phi\pm(\beta^2-\alpha^2)\cos2\phi\Big)\Big], 
\label{eq:4}
\end{align}

and for the magnetic contribution

\begin{align}
s_{z}^\text{B} &= \frac{\epsilon_0 k_{t}^2}{2k^2\omega}\Big[J_{|\ell|}J_{|\ell|-2}\Big(\mp2\alpha\beta\sin2\phi\pm(\beta^2-\alpha^2)\cos2\phi\Big) \nonumber \\ & + J_{|\ell|}J_{|\ell|+2}\Big(\pm2\alpha\beta\sin2\phi\pm(\alpha^2-\beta^2)\cos2\phi\Big)\Big]. 
\label{eq:5}
\end{align}

\begin{figure} []
\includegraphics[width=0.55\textwidth]{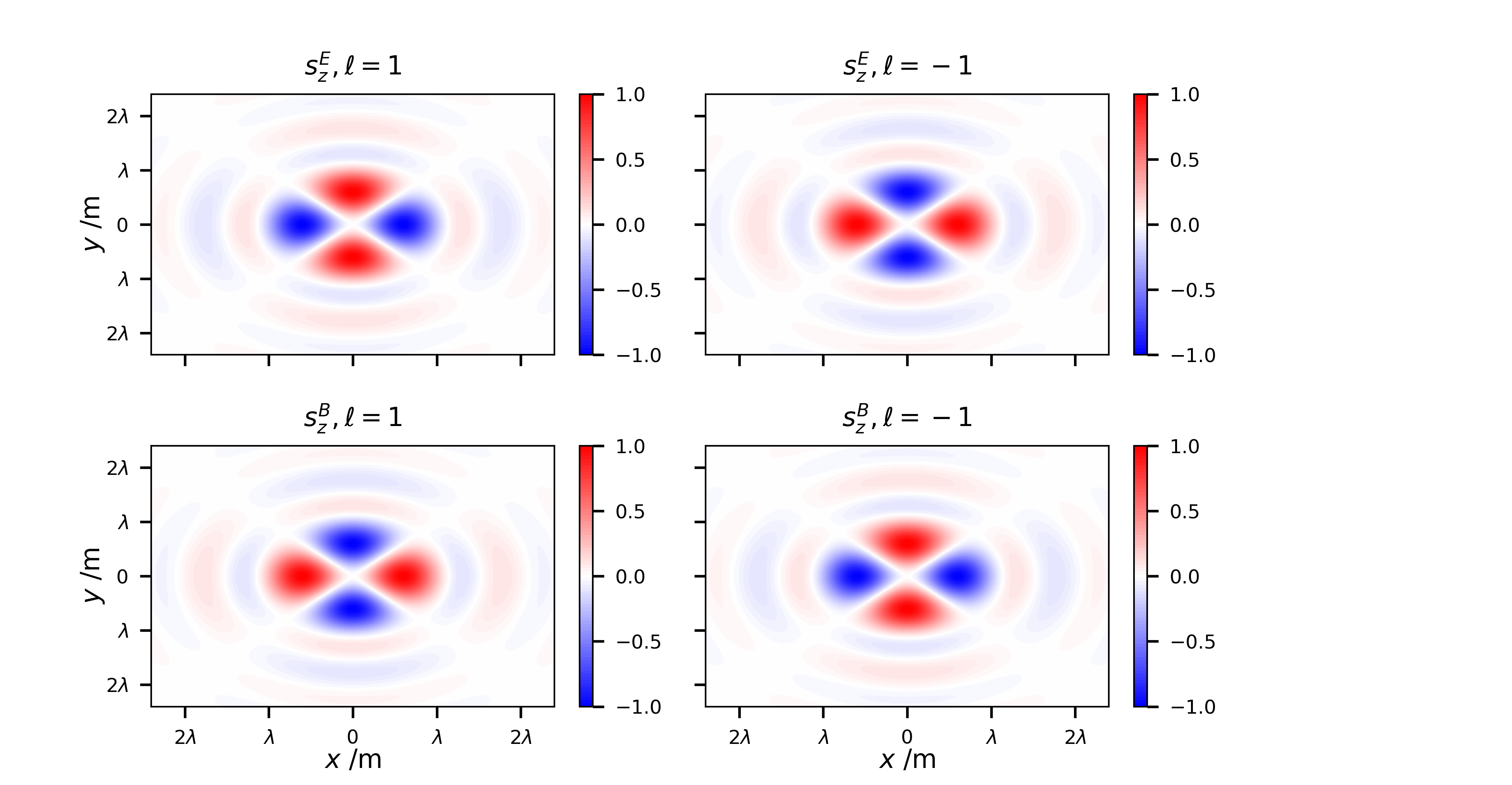}
\caption{Normalizd electric $s_z^E$ Eq.~\eqref{eq:4} and magnetic $s_z^B$ Eq.~\eqref{eq:5} longitudinal spin angular momentum density at $z=0$ (the focal plane) for tightly-focused $k_t/k_z = 0.6315$ 2D $x$-polarized Bessel beams (i.e. $\alpha = 1, \beta = 0$) with topological charge $\ell = \pm1$. Clearly $s_z = s_z^E + s_z^B = 0$ and the sign of the topological charge flips the sign of the spin momentum density locally. Although not shown, it is clear from Eqs.~\eqref{eq:4} and \eqref{eq:5} the spatial distribution of the local spin density rotates with the azimuth of the 2D linear polarization state of the input beam, e.g. the above plots would be rotated by $\pi/2$ for 2D $y$-polarized input Bessel beams. $\lambda = 729$ nm in all plots.}
    \label{fig:1}
\end{figure}

For 2D linearly polarized input beams the $\mathbf{E}_{\text{T}0} \times \mathbf{E}_{\text{T}0}$ and $\mathbf{B}_{\text{T}0} \times \mathbf{B}_{\text{T}0}$ contributions to $s_z$ are obviously zero. As stated, the non-zero longitudinal spin densities Eqs.~\eqref{eq:4} and \eqref{eq:5} are generated from the cross terms $\mathbf{E}_{\text{T}0} \times \mathbf{E}_{\text{T}2}$ and $\mathbf{B}_{\text{T}0} \cross \mathbf{B}_{\text{T}2}$, respectively. It is clear to see that $s_z^\text{E} = - s_z^\text{B}$ and $s_z = s_z^\text{E} + s_z^\text{B} = 0$. Nonetheless, this does not preclude the experimental observation of either of these spin angular momentum densities due to the dual-asymmetric nature of most materials. For an electric (magnetic) dipole particle a torque would be generated through $s_z^\text{E}$ ($s_z^\text{B}$). The spatial distributions of Eqs.~\eqref{eq:4} and \eqref{eq:5} are given in Fig.~\ref{fig:1} for $\alpha = 1$. What is precluded is a chiral force on a  dipolar chiral particle stemming from a dissapative, non-conservative radiation pressure due to this total spin of the field \cite{bliokh2014magnetoelectric, hayat2015lateral, genet2022chiral}, i.e. $\mathbf{F}\propto\text{Im}\chi s_z \hat{\mathbf{z}}=0$, where $\chi$ is the chiral dipolar polarizability (see Discussion for more on this).

\section{Helicity density}

We now highlight that the longitudinal spin angular momentum density of the previous section has no association with a corresponding optical helicity density. The dual symmetric canonical momentum density for a monochromatic cycle averaged electromagnetic field is

\begin{align}
\mathbf{p}_\text{o} = \frac{\epsilon_0}{4\omega} \text{Im}\Big[\mathbf{E}^*\cdot\nabla \mathbf{E} + c^2\mathbf{B}^*\cdot\nabla \mathbf{B}\Big]. 
\label{eq:6}
\end{align}

We require the canonical momentum density which stems purely from the zeroth-order transverse fields in our analysis. This is easily calculated:

\begin{align}
\mathbf{p}_{\text{o}}^\text{E} &= \frac{\epsilon_0}{4\omega} J_{|\ell|}^2\Big(\frac{\ell}{r}\hat{\phi} + k_z \hat{z}\Big), 
\label{eq:7} \\
\mathbf{p}_{\text{o}}^\text{B} &= \frac{\epsilon_0}{4\omega}\frac{k{_z}^2}{k^2} J_{|\ell|}^2\Big(\frac{\ell}{r}\hat{\phi} + k_z \hat{z}\Big). 
\label{eq:8}
\end{align}

The helicity density $h$ may be defined as the projection of the spin density on to the linear momentum density, i.e. $h = \mathbf{s}\cdot\mathbf{p}_\text{o}/p_\text{o}(k_z)$ \cite{bliokh2015transverse}. With this definition we can calculate the following 'electric' and 'magnetic' contributions of the helicity associated with the spin momentum densities Eqs.~\eqref{eq:4} and \eqref{eq:5}:

\begin{align}
h^\text{E} &= \mathbf{s}^\text{E}\cdot\mathbf{p}_{\text{o}}^\text{E}/p_{\text{o}}^\text{E}(k_z) \nonumber \\ &= \Big[J_{|\ell|}J_{|\ell|-2}\Big(\pm2\alpha\beta\sin2\phi\pm(|\alpha|^2-|\beta|^2)\cos2\phi\Big) \nonumber \\ & + J_{|\ell|}J_{|\ell|+2}\Big(\mp2\alpha\beta\sin2\phi\pm(|\beta|^2-|\alpha|^2)\cos2\phi\Big)\Big], 
\label{eq:9}
\end{align}

and 

\begin{align}
h^\text{B} &= \mathbf{s}^\text{B}\cdot\mathbf{p}_{\text{o}}^\text{B}/p_{\text{o}}^\text{B}(k_z) \nonumber \\ &= \Big[J_{|\ell|}J_{|\ell|-2}\Big(\mp2\alpha\beta\sin2\phi\pm(|\beta|^2-|\alpha|^2)\cos2\phi\Big) \nonumber \\ & + J_{|\ell|}J_{|\ell|+2}\Big(\pm2\alpha\beta\sin2\phi\pm(|\alpha|^2-|\beta|^2)\cos2\phi\Big)\Big],
\label{eq:10}
\end{align}

Which leads to the fact that:

\begin{align}
h = h^\text{E} + h^\text{B} = 0.
\label{eq:11}
\end{align}

Thus there is no helicity associated with the spin momentum density of 2D linearly polarized optical vortices even though it is longitudinal with respect to the direction of beam propagation. Although there are `individual' non-zero electric and magnetic contributions to the helicity, the optical helicity is a dual symmetric quantity by its very nature \cite{bliokh2011characterizing, aiello2022helicity} and couples to the dual-symmetric chiral polarizability of particles which mixes electric and magnetic dipoles. Thus one cannot experimentally determine the individual contributions $h^\text{E}$ and $h^\text{B}$ (unlike spin density $\mathbf{s}^\text{E}$, $\mathbf{s}^\text{B}$, energy density $w^\text{E}$, $w^\text{B}$ , canonical momentum density $\mathbf{p}_\text{o}^\text{E}$, $\mathbf{p}_\text{o}^\text{B}$, etc.). These results can be verified using the more standard definition of the optical helicity density of a monochromatic beam and using Eqs.~\eqref{eq:1} and \eqref{eq:2}: 

\begin{align}
h &= - \frac{\epsilon_0 \omega}{2}\text{Im}\mathbf{E}^*\cdot\mathbf{B} \nonumber \\
& = - \frac{\epsilon_0 \omega}{2}\text{Im}(\mathbf{E}_{\text{T}_0}^*+\mathbf{E}_{\text{T}_2}^*)\cdot(\mathbf{B}_{\text{T}_0}+\mathbf{B}_{\text{T}_2}) = 0, 
\label{eq:12}
\end{align}

because $\mathbf{E}_{\text{T}_0}^*\cdot\mathbf{B}_{\text{T}_0} = 0$, $\mathbf{E}_{\text{T}_2}^*\cdot\mathbf{B}_{\text{T}_2} = 0$, and importantly $\mathbf{E}_{\text{T}_0}^*\cdot\mathbf{B}_{\text{T}_2} = - \mathbf{E}_{\text{T}_2}^*\cdot\mathbf{B}_{\text{T}_0}$ \cite{forbes2021measures}. In the above, the $\mathbf{E}_{\text{T}_0}^*\cdot\mathbf{B}_{\text{T}_2}$ term is associated with $h^\text{B}$ and $\mathbf{E}_{\text{T}_2}^*\cdot\mathbf{B}_{\text{T}_0}$ with $h^\text{E}$. We have neglected the $\mathbf{E}_{\text{L}_1}^*\cdot\mathbf{B}_{\text{L}_1}$ contribution to the helicity in Eq.~\eqref{eq:12} because for 2D linearly polarized input beams it is associated with the transverse spin momentum density (see Discussion for further). 

\section{Discussion and Conclusion}

Clearly then electromagnetic fields can carry a measurable electric Eq.~\eqref{eq:4} and magnetic Eq.~\eqref{eq:5} longitudinal spin angular momentum density even though they have no associated helicity Eqs.~\eqref{eq:9}-\eqref{eq:12}. It is therefore apparent that it is not correct to equate helicity with longitudinal spin. If one inspects what the equations for spin density and helicity density actually quantify in monochromatic beams this becomes readily apparent. The spin density Eq.~\eqref{eq:3} measures the degree to which two orthogonal field components (electric or magnetic) are $\pi/2$ out of phase: it measures the degree of polarization ellipticity in a given 2D plane. The local spin density of light is thus correlated to polarization ellipticity. Looking at Eqs.~\eqref{eq:1} and \eqref{eq:2} it is clear that the $\text{T}_0$ and $\text{T}_2$ fields possess components with a $\pi/2$ out-of-phase relationship. For example, if $\alpha = 1$ then the $x$-polarized $\text{T}_0$ component in Eq.~\ref{eq:1} and the $y$-polarized $\text{T}_2$ components are $\pi/2$ out of phase due to the factor of $i$ in the latter, thus in the $x,y$ plane the electric field vector traces out an elliptical (in general) path generating the local electric spin angular momentum. Under paraxial conditions $k_t<<k_z$ and the factor $k_t^2/k^2$ is minuscule leading to an essentially non-existent $\text{T}_2$ field. The paraxial beam is thus essentially just $x$-polarized in the $x,y$ plane due to the large $\text{T}_0$ field; as the beam becomes more focused the ratio of $k_t/k_z$ grows and the polarization vector becomes elliptical in the $x,y$-plane due to the $\pi/2$ out of phase $\text{T}_2$ $y$-component.  

On the other hand, the optical helicity Eq.~\eqref{eq:12} is a measure of whether parallel electric and magnetic field components (in a single direction $x$, $y$, or $z$) are $\pi/2$ out of phase, which is in general completely independent of polarization ellipticity (and thus spin). This is why the longitudinal fields of optical vortex beams, both polarized in $z$ and thus cannot have ellipticity, can generate non-zero optical helicity \cite{forbes2021measures, forbes2022optical, green2023optical, forbes2023customized} via $\mathbf{E}_{\text{L}_1}^*\cdot\mathbf{B}_{\text{L}_1}$. Although we have shown there is no helicity associated with the longitudinal spin momentum densities Eqs.~\eqref{eq:4} and \eqref{eq:5}, remarkably the polarization-independent helicity density of optical vortex beams is associated with a transverse spin momentum density \cite{forbes2021measures}. This is due to the fact that the canonical momentum density of an optical vortex has an azimuthal component in addition to the standard longitudinal component (see Eqs.~ \eqref{eq:7} and \eqref{eq:8}). When the transverse spin momentum density of a vortex beam is projected on to the transverse linear momentum density it produces the non-zero polarization independent helicity of vortex beams. While transverse spin is a generic property of spatially confined electromagnetic fields, an azimuthal component of linear momentum density is unique to optical vortex beams and this accounts for their extraordinary optical helicity properties. 

Finally it is worth briefly revisiting the fact that the longitudinal spin of tightly-focused 2D linearly polarized vortices can not yield a chiral radiation pressure force $\mathbf{F}\propto\text{Im}\chi s_z \hat{\mathbf{z}}=0$. As we have now established, there is no helicity associated with this spin angular momentum, and thus no pseudoscalar quantity which is required for a true chiral light-matter interaction. Note this does not preclude or negate the chiral radiation pressure force which can manifest through the curl of the Poynting vector \cite{bliokh2014magnetoelectric, genet2022chiral}.

In conclusion we have identified that the longitudinal spin density of electromagnetic fields can exist without an associated helicity density; previous work has shown that  helicity density can be associated with a transverse spin density. These remarkable conclusions manifest most clearly in non-paraxial (tightly focused) optical vortex beams. Such behaviour is in stark contrast to the canonical picture of spin and helicity, which applies to paraxial beams and plane waves.   

\section{Appendix: transverse spin momentum density}

The transverse spin momentum density components for a 2D linearly-polarized Bessel beam are calculated to be

\begin{align}
s_{x}^\text{E} &= \frac{\epsilon_0 k_{t}}{2k_z\omega}\Big[J_{|\ell|}J_{|\ell|-1}\Big(\alpha\beta\cos\phi+\beta^2\sin\phi\Big) \nonumber \\ & - J_{|\ell|}J_{|\ell|+1}\Big(\alpha\beta\cos\phi+\beta^2\sin\phi\Big) \Big], 
\label{eq:14}
\end{align}

\begin{align}
s_{y}^\text{E} &= -\frac{\epsilon_0 k_{t}}{2k_z\omega}\Big[J_{|\ell|}J_{|\ell|-1}\Big(\alpha\beta\sin\phi+\alpha^2\cos\phi\Big) \nonumber \\ & - J_{|\ell|}J_{|\ell|+1}\Big(\alpha\beta\sin\phi+\alpha^2\cos\phi\Big) \Big], 
\label{eq:15}
\end{align}

\begin{align}
s_{x}^\text{B} &= \frac{\epsilon_0 k_{t} k_z}{2k^2\omega}\Big[J_{|\ell|}J_{|\ell|-1}\Big(\alpha^2\sin\phi-\alpha\beta\cos\phi\Big) \nonumber \\ & - J_{|\ell|}J_{|\ell|+1}\Big(\alpha^2\sin\phi-\alpha\beta\cos\phi\Big) \Big], 
\label{eq:16}
\end{align}

and

\begin{align}
s_{y}^\text{B} &= \frac{\epsilon_0 k_{t} k_z}{2k^2\omega}\Big[J_{|\ell|}J_{|\ell|-1}\Big(\alpha\beta\sin\phi-\beta^2\cos\phi\Big) \nonumber \\ & + J_{|\ell|}J_{|\ell|+1}\Big(\beta^2\cos\phi -\alpha\beta\sin\phi+\Big) \Big]. 
\label{eq:17}
\end{align}

These terms a generated via: $s_{x,y}^\text{E} = \mathbf{E}_{\text{T}0} \times \mathbf{E}_{\text{L}1}$ and $s_{x,y}^\text{B} = \mathbf{B}_{\text{T}0} \times \mathbf{B}_{\text{L}1}$, and projected on to the transverse canonical momentum density yield a 2D polarization-independent optical helicity density.

\bibliographystyle{apsrev4-1}
\bibliography{references.bib}
\end{document}